\documentclass[12pt,prd,showpacs,tightenlines,nofootinbib]{revtex4}
\usepackage{bm}
\usepackage{graphics}
\usepackage{rotating}
\usepackage{epsfig}
\begin{document}
\title{Exclusive weak $B$ decays involving $\tau$ lepton in the relativistic
  quark model} 
\author{R. N. Faustov}
\author{V. O. Galkin}
\affiliation{Dorodnicyn Computing Centre, Russian Academy of Sciences,
  Vavilov Str. 40, 119991 Moscow, Russia}

\begin{abstract}
Semileptonic and leptonic $B$ decays are analyzed in the framework of
the relativistic quark model. Special attention is payed to the decays
involving $\tau$ lepton. It is found that the calculated particular
decay branching fractions are consistent with available experimental
data within error bars. However, the predicted and recently measured
ratios $R(D^{(*)})$ of the $B\to D^{(*)}\tau\nu_\tau$ and $B\to
D^{(*)}l\nu_l$ branching fractions differ by 1.75$\sigma$ for
$R(D)$ and by 2.4$\sigma$ for $R(D^*)$.          
\end{abstract}
\pacs{13.20.He, 12.39.Ki}

\maketitle

Recently significant progress has been achieved in studying the exclusive
semileptonic and weak leptonic $B$ decays. Branching fractions of
these decays involving $\tau$ lepton were measured
\cite{babar,belle,babarR,pdg,belletau,babartau}. It was found that the experimental values of
the semileptonic $B\to D^{(*)}\tau\nu_\tau$ \cite{babar,belle,babarR,pdg} and leptonic $B\to
\tau\nu_\tau$ \cite{pdg,belletau,babartau} decays differ by more than $2\sigma$ from the
theoretical predictions in the framework of the Standard Model. This
deviation is mostly pronounced when theoretical expectations are
compared with the recent BaBar data for the ratios $R(D^{(*)})$ of the semileptonic
$B\to D^{(*)}\tau\nu_\tau$ and  $B\to  D^{(*)}l\nu_l$ decay branching fractions \cite{babarR}. In such
ratios most of the uncertainties, e.g. the ones emerging
from the determination of the Cabibbo-Kobayashi-Maskawa (CKM) matrix
element $V_{cb}$, cancel. However, such ratios of decay rates still
depend on meson form factors, and thus are model dependent, but this
dependence is significantly milder than for the individual branching
fractions. These measurements initiated discussion in the literature
on the possible New Physics contributions to the decay processes
involving $\tau$ lepton (see e.g. \cite{nph} and references therein).    

In this letter we give predictions for these weak semileptonic and
leptonic $B$ decays in the framework of the relativistic quark model.
The model is based on the
quasipotential approach in quantum field theory with the QCD motivated
interaction.  Hadrons are considered as the bound states of
constituent quarks and are described by the
single-time wave functions satisfying the
three-dimensional Schr\"odinger-like equation, which is
relativistically invariant \cite{bcprop}: 
\begin{equation}
\label{quas}
{\left(\frac{b^2(M)}{2\mu_{R}}-\frac{{\bf
p}^2}{2\mu_{R}}\right)\Psi_{M}({\bf p})} =\int\frac{d^3 q}{(2\pi)^3}
 V({\bf p,q};M)\Psi_{M}({\bf q}),
\end{equation}
where the relativistic reduced mass is
\begin{equation}
\mu_{R}=\frac{M^4-(m^2_1-m^2_2)^2}{4M^3},
\end{equation}
 $M$ is the meson mass, $m_{1,2}$ are the quark masses,
and ${\bf p}$ is their relative momentum.  
In the center of mass system the relative momentum squared on mass shell 
reads
\begin{equation}
{b^2(M) }
=\frac{[M^2-(m_1+m_2)^2][M^2-(m_1-m_2)^2]}{4M^2}.
\end{equation}
The interaction quasipotential $V({\bf p,q};M)$ consists of the perturbative one-gluon
exchange part and the nonperturbative confining part \cite{bcprop}. The Lorentz
structure of the latter part includes the scalar and vector linearly
rising interactions. The long-range vector vertex contains the Pauli
term (anomalous chromomagnetic quark moment) which
enables vanishing of the spin-dependent chromomagnetic interaction in
accord with the flux tube model. 

\begin{figure}
  \centering
 \includegraphics[width=8cm]{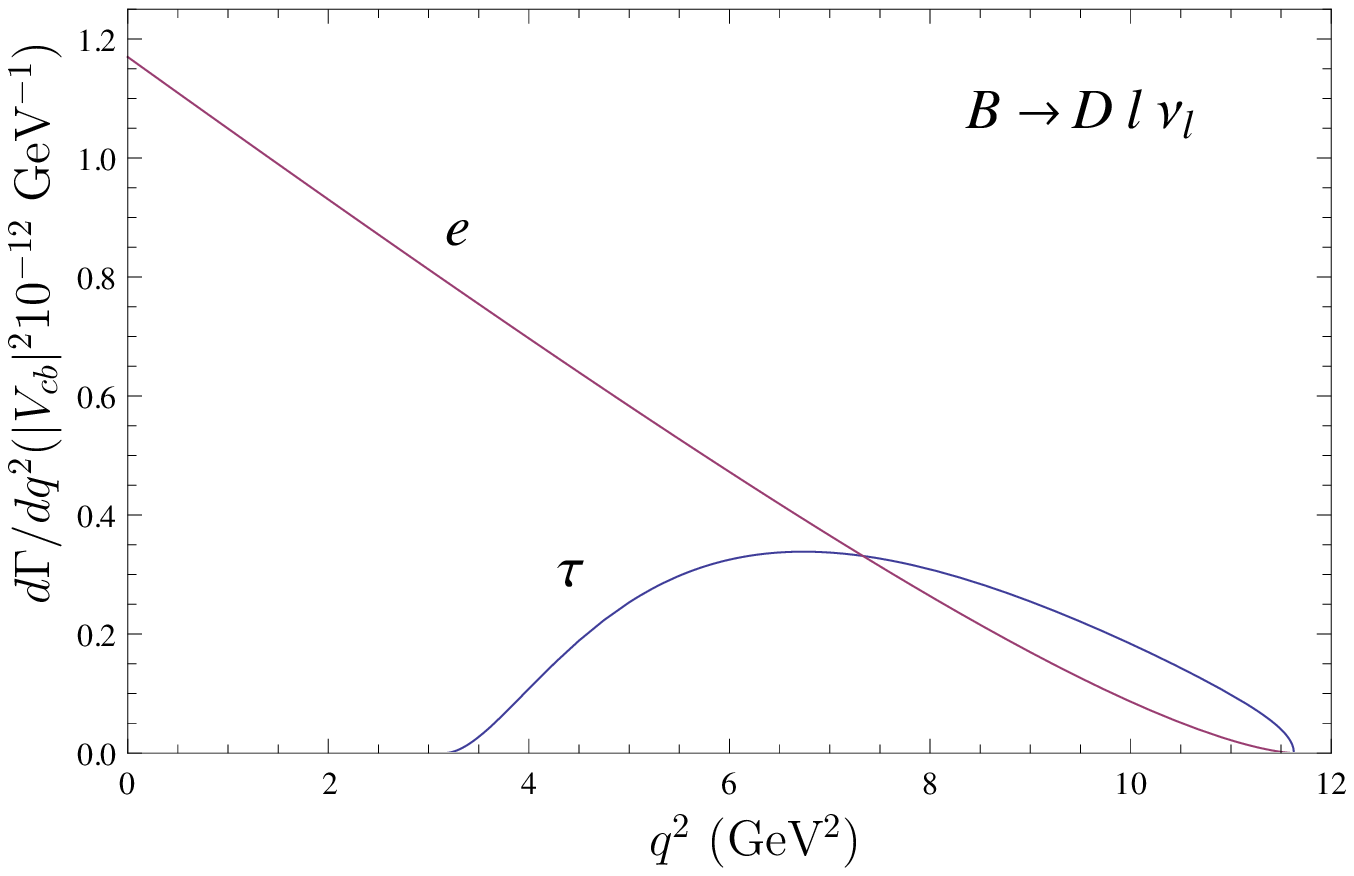}\ \
 \  \includegraphics[width=8cm]{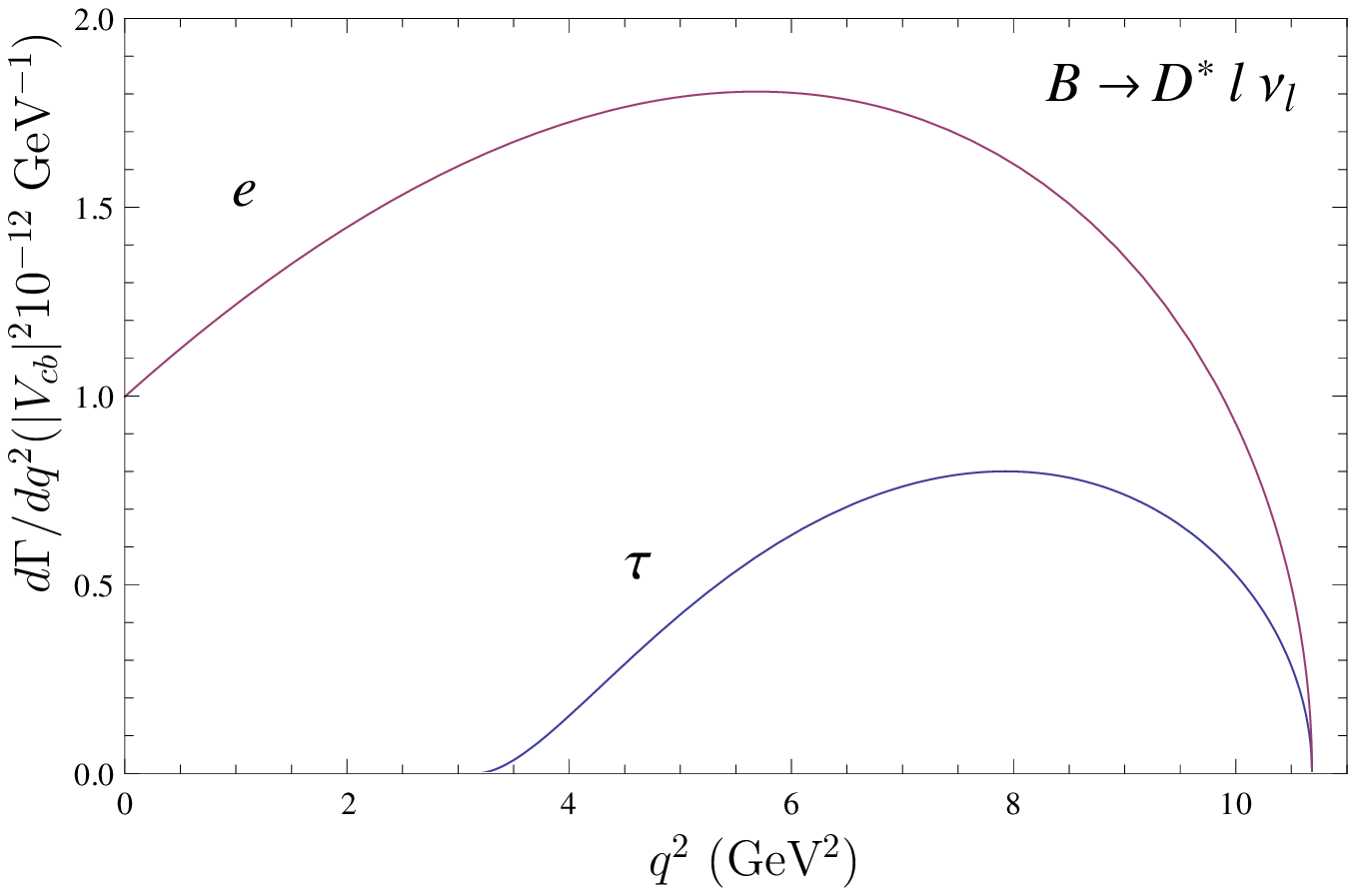}

  \caption{Predictions for the differential decay rates  of the $B\to D^{(*)}$ 
    semileptonic decays. }
  \label{fig:brbd}
\end{figure}

The semileptonic $B$ decay form factors were calculated in our model
in Ref.~\cite{asbd}. For the heavy-to-heavy $B\to D^{(*)}$ transitions
the heavy quark expansion has been employed. Leading and subleading
order Isgur-Wise functions were explicitly determined as overlap
integrals of the meson wave functions and on this basis the decay
form factors were calculated in the whole kinematical range up to
$1/m_Q$ terms. These form factors 
and decay branching fractions  were found to agree well with the
available experimental data. We calculated  decay rates involving 
light leptons $l=\mu,\nu$ since at that time only such decays were
measured experimentally. Now we apply these form factors for the
consideration of the $B\to  D^{(*)}\tau\nu_\tau$ decays. The obtained
branching fractions for them as well as updated predictions for $B\to
D^{(*)}l\nu_l$ decays are given in Table~\ref{tab:br} in comparison
with experimental data. Both experimental averages form PDG
\cite{pdg} and recent BaBar data \cite{babarR} are given. For
theoretical estimates we use the CKM matrix element
$|V_{cb}|=0.039\pm0.0015$, which is obtained from the
comparison of our theoretical predictions \cite{asbd} for the products
$F_{D^{(*)}}(w)|V_{cb}|$ and for the $B\to D^{(*)}l\nu_l$ decay branching
fractions with updated experimental data.~\footnote{This value of $|V_{cb}|$  is in
  accord with its recent evaluation by the Heavy Flavor
  Averaging Group \cite{hfag}.}  The uncertainties in our
theoretical predictions originate mainly from the $|V_{cb}|$ value and higher order
$1/m_Q$ contributions to form factors. We find that almost all measured and
calculated values of branching fractions agree within
uncertainties. This is also true for the decays involving 
$\tau$ lepton. In Fig.~\ref{fig:brbd} we show our predictions for the
differential decay rates  of the $B\to D^{(*)}l\nu_l$  and $B\to
D^{(*)}\tau\nu_\tau$ semileptonic decays.       

\begin{table}
\caption{Calculated ($|V_{cb}|=0.039\pm0.0015$) and measured branching fractions of semileptonic $B\to
  D^{(*)}l\nu_{l}$ and $B\to
  D^{(*)}\tau\nu_{\tau}$ decays (in \%).} 
\label{tab:br}
\begin{ruledtabular}
\begin{tabular}{cccc}
Decay & Theory & Experiment \cite{pdg} & Experiment \cite{babarR} \\
\hline
$B^0\to D^-l^+\nu_l$ & $2.14\pm0.16$ &$2.18\pm0.12$
&$2.23\pm0.11\pm0.11$\\ 
$B^0\to D^-\tau^+\nu_\tau$ & $0.68\pm0.05$ &$1.1\pm0.4$
&$1.01\pm0.18\pm0.12$\\
$B^+\to D^0l^+\nu_l$ & $2.32\pm0.17$ &$2.26\pm0.11$
&$2.31\pm0.08\pm0.09$\\ 
$B^+\to D^0\tau^+\nu_\tau$ & $0.73\pm0.05$ &$0.77\pm0.25$
&$0.99\pm0.19\pm0.13$\\
$B^0\to D^{*-}l^+\nu_l$ & $5.51\pm0.42$ &$4.95\pm0.11$
&$4.72\pm0.05\pm0.34$\\ 
$B^0\to D^{*-}\tau^+\nu_\tau$ & $1.43\pm0.11$ &$1.5\pm0.5$
&$1.74\pm0.19\pm0.12$\\
$B^+\to D^{*0}l^+\nu_l$ & $6.00\pm0.46$ &$5.70\pm0.19$
&$5.40\pm0.02\pm0.21$\\ 
$B^+\to D^{*0}\tau^+\nu_\tau$ & $1.56\pm0.12$ &$2.04\pm0.30$ &$1.71\pm0.17\pm0.13$\\
\end{tabular}
\end{ruledtabular}
\end{table}

\begin{figure}
  \centering
 \includegraphics[width=8cm]{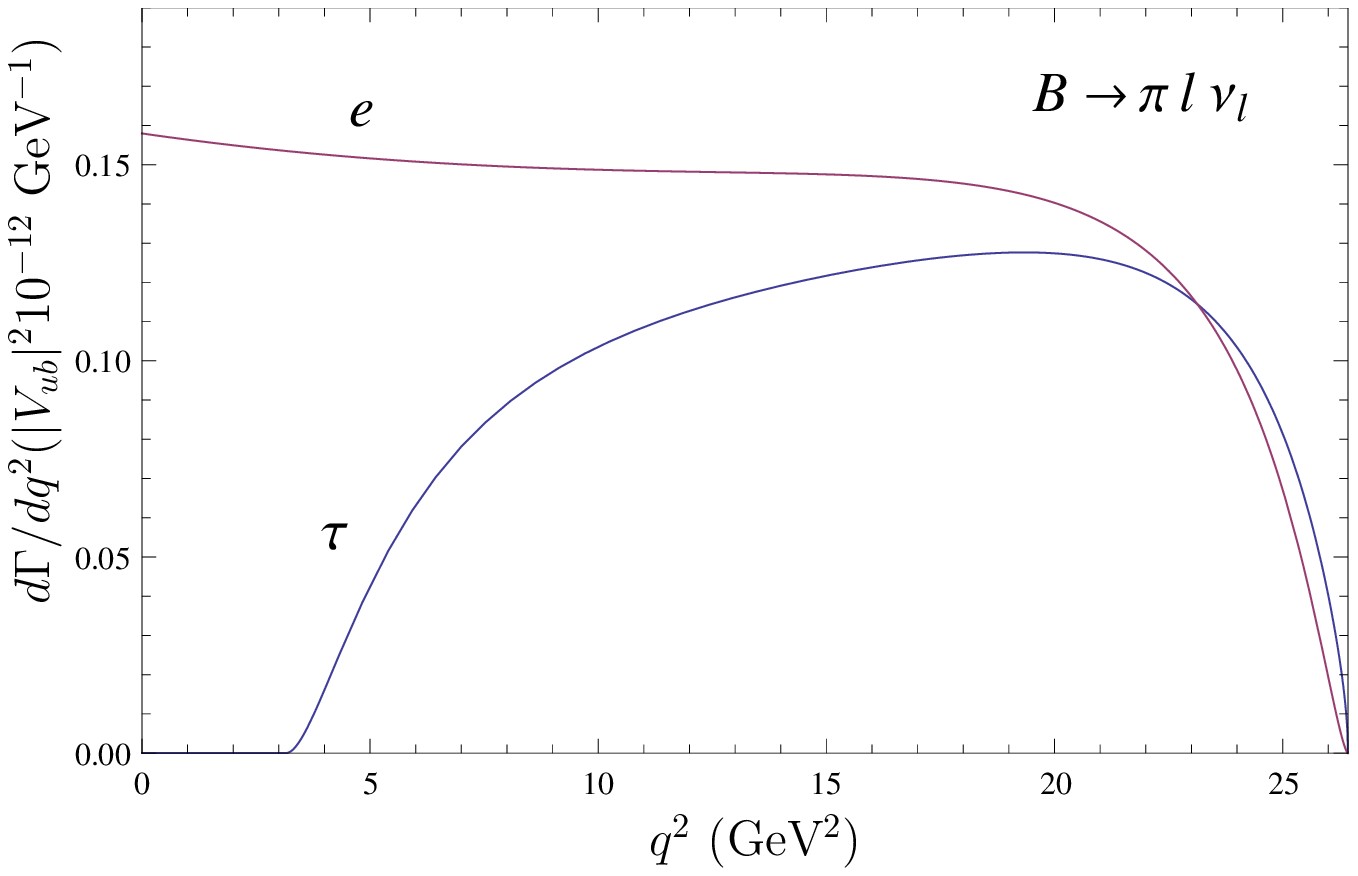}\ \
 \  \includegraphics[width=8cm]{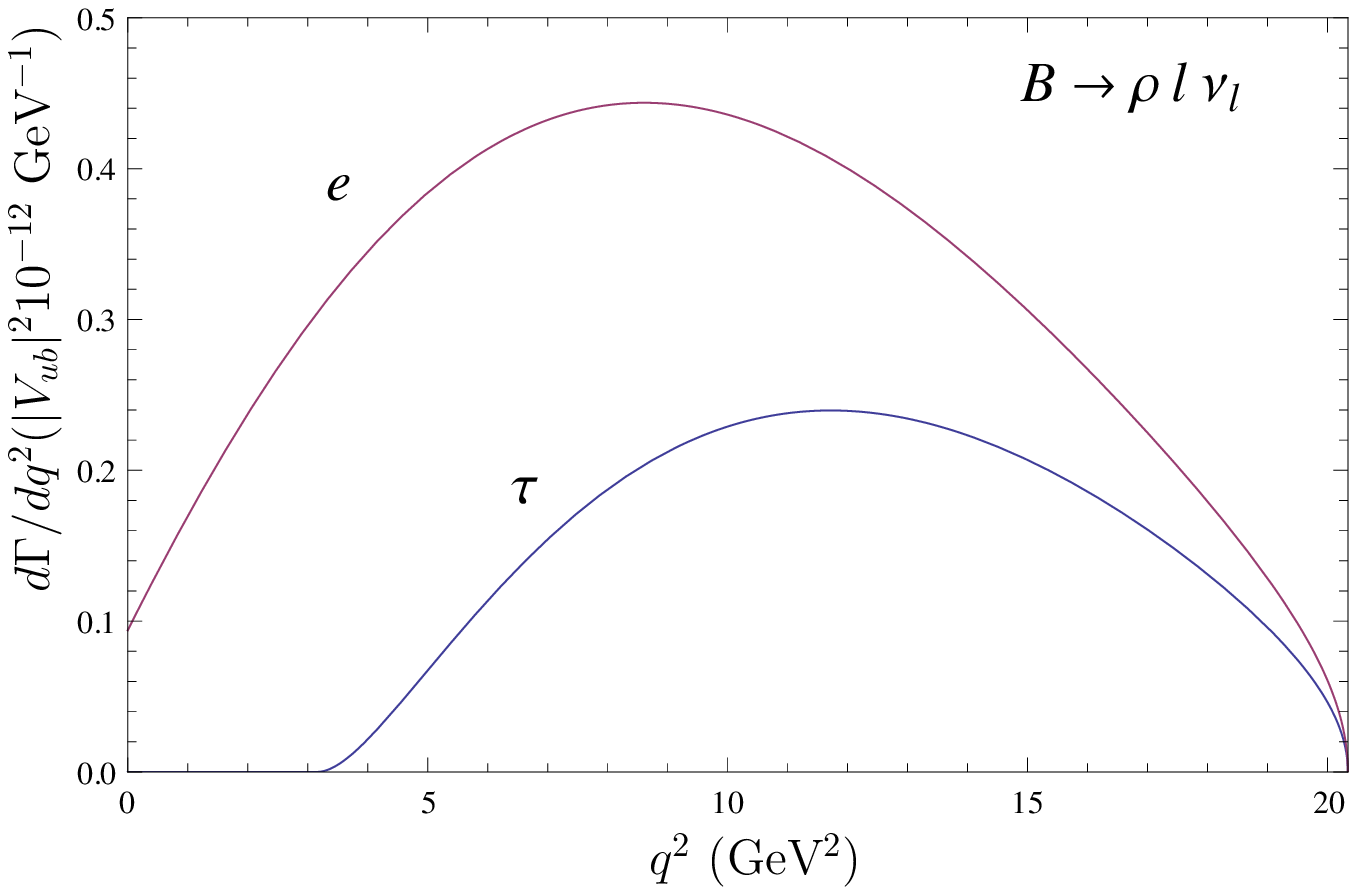}

  \caption{Predictions for the differential decay rates  of the $B\to \pi(\rho)$ 
    semileptonic decays. }
  \label{fig:brbpr}
\end{figure}

Form factors of the heavy-to-light semileptonic $B$ decays, such as
$B\to \pi(\rho)l\nu_l$, have been calculated in our model in Ref.~
\cite{asbd} without application of the heavy quark expansion. They
were expressed through the overlap integrals of the meson wave
functions which are known form the meson mass spectra calculations. The
momentum transfer dependence of these form factors was determined in
the whole kinematical range without any ad hoc parametrizations.
Relativistic effects were systematically taken into account, including
negative-energy contributions and relativistic transformation of the
meson wave function from the rest to the moving reference frame. In
Tables~\ref{tab:dtn}, \ref{tab:brpi} we present results of the updated calculation of
the semileptonic $B\to \pi(\rho)l\nu_{l}$  decay branching fractions
in comparison with experimental data \cite{pdg,hfag}. We also give
predictions for the heavy-to-light semileptonic decays involving
$\tau$ lepton $B\to \pi(\rho)\tau\nu_{\tau}$. Corresponding
plots for the semileptonic differential decay rates are shown
in Fig.~\ref{fig:brbpr}. The value of the CKM matrix element $|V_{ub}|$
obtained from the comparison of our predictions and measured $B\to
\pi(\rho)l\nu_{l}$ decay rates amounts
$|V_{ub}|=(4.05\pm0.20)\times10^{-3}$. As we see from
Tables~\ref{tab:dtn}, \ref{tab:brpi} calculated decay branching fractions agree
well with available experimental data.

\begin{table}
\caption{Calculated [$|V_{ub}|=(4.05\pm0.20)\times10^{-3}$] and
  measured branching fractions of the semileptonic $B\to
  \pi(\rho)l\nu_{l}$ and $B\to \pi(\rho)
 \tau\nu_{\tau}$ decays ($\times10^{-4}$).} 
\label{tab:dtn}
\begin{ruledtabular}
\begin{tabular}{ccc}
Decay & Theory& Experiment \cite{pdg}  \\
\hline
$B^0\to\pi^-l^+\nu_{l}$ & $1.37\pm0.13$ &$1.44\pm0.05$\\
$B^0\to\pi^-\tau^+\nu_{\tau}$ & $0.86\pm0.08$ &\\
$B^+\to\pi^0l^+\nu_{l}$ & $0.74\pm0.07$ &$0.778\pm0.028$\\
$B^+\to\pi^0\tau^+\nu_{\tau}$ & $0.47\pm0.05$ &\\
$B^0\to\rho^-l^+\nu_{l}$ & $2.40\pm0.24$ &$2.34\pm0.28$\\
$B^0\to\rho^-\tau^+\nu_{\tau}$ & $1.04\pm0.10$ &\\
$B^+\to\rho^0l^+\nu_{l}$ & $1.29\pm0.13$ &$1.07\pm0.13$\\
$B^+\to\rho^0\tau^+\nu_{\tau}$ & $0.56\pm0.06$ &\\
\end{tabular}
\end{ruledtabular}
\end{table}

\begin{table}
\caption{ Comparison of theoretical predictions and experimental
  averages for exclusive determinations of  $Br(\bar B\to
  \pi l\bar\nu)$ ($\times10^{-4}$).} 
\label{tab:brpi}
\begin{ruledtabular}
\begin{tabular}{ccccc}
& $Br$ & $Br(q^2<12~{\rm GeV}^2)$ &$Br(q^2<16~{\rm GeV}^2)$&$Br(q^2>16~{\rm GeV}^2)$ \\
\hline
Theory & $1.43\pm 0.14$ & $0.72\pm0.07$ & $0.96\pm0.09$ &
$0.47\pm0.05$\\
Experiment \cite{hfag} &$1.42\pm0.03\pm0.04$&$0.81\pm0.02\pm0.02$
&$1.05\pm0.02\pm0.03$ &$0.37\pm0.01\pm0.02$
\end{tabular}
\end{ruledtabular}
\end{table}

Next we consider the ratios of $B$ decays involving light $l$ and
$\tau$ leptons
\begin{equation}
  \label{eq:r}
  R(M)=\frac{Br(B\to M\tau\nu_\tau)}{Br(B\to Ml\nu_l)}, \qquad M=D^{(*)},\pi,\rho.
\end{equation}
In these ratios the corresponding CKM matrix elements cancel and as a
result uncertainties are significantly reduced. We present our
predictions for $R(M)$ in Table~\ref{tab:rbf} and confront them with
recent experimental values for $R(D)$ and $R(D^*)$ from BaBar
Collaboration \cite{babarR}. It is important to note that most of the
above mentioned form factor uncertainties cancel out in these
ratios. We see that our predictions for these ratios are lower than
the experimental values by 1.75$\sigma$ for $R(D)$ and by 2.4$\sigma$ for
$R(D^*)$.

\begin{table}
\caption{Ratios of branching fractions of the semileptonic $B\to
  M\tau\nu_\tau$ and $B\to
  Ml\nu_l$ ($M=D^{(*)},\pi,\rho$) decays.} 
\label{tab:rbf}
\begin{ruledtabular}
\begin{tabular}{ccc}
Ratio & Theory & Experiment \cite{babarR} \\
\hline
$R(D)$ & 0.315 
&$0.440\pm0.058\pm0.042$\\ 
$R(D^*)$ & 0.260 
&$0.332\pm0.024\pm0.018$\\
$R(\pi)$& 0.63\\
$R(\rho)$& 0.43
\end{tabular}
\end{ruledtabular}
\end{table}

We use the obtained above value of the CKM matrix element
$|V_{ub}|=(4.05\pm0.20)\times10^{-3}$ to calculate the branching
fractions of the leptonic $B\to L\nu_L$ ($L=l,\tau$) decays:
\begin{equation}
  Br(B\to L\nu_L)=\frac{G_F^2 }{8 \pi}M_B M_L^2 \left(1 - \frac{M_L^2}{M_B^2}\right)^2 f_B^2 |V_{ub}|^2 \tau_B, 
\end{equation}
where $M_B$ and $M_L$ are the $B$ meson and $L$ lepton masses, $G_F$ is
the Fermi constant, $f_B$ is the decay constant and $\tau_B$
is the  life time of the $B$ meson. The decay constants of light and
heavy-light mesons were calculated in our model in Ref.~\cite{dconst}
with the consistent account of relativistic effects including contributions
of the negative-energy quark states. The obtained value of the $B$
meson decay constant $f_B=(189\pm9)$~MeV is in good agreement with the
unquenched lattice QCD calculations $f_B^{\rm lattice} = (189 \pm
4)$~MeV \cite{fbl}. The result for 
the branching fraction of the $B\to\tau\nu_{\tau}$ decay  is
compared to the experimental data in Table~\ref{tab:dbl}, where
the PDG \cite{pdg} as well as new Belle \cite{belletau} and
BaBar \cite{babartau} data are given. We find that our prediction for
this decay branching fraction is lower than central experimental
values. However, taking into account large uncertainties, it agrees
well with separate Belle and BaBar measurements, but 
deviates by 1.25$\sigma$ from the averaged value \cite{pdg}.

\begin{table}
\caption{Calculated [$f_B=(189\pm9)$~MeV, $f_{B_c}=(433\pm5)$~MeV, $|V_{ub}|=(4.05\pm0.20)\times10^{-3}$, $|V_{cb}|=0.039\pm0.0015$] and measured branching fractions of the leptonic $B$ and $B_c$  decays ($\times10^{-4}$).} 
\label{tab:dbl}
\begin{ruledtabular}
\begin{tabular}{ccccc}
Decay & Theory& Experiment \cite{pdg} & Experiment \cite{belletau} & Experiment \cite{babartau} \\
\hline
$B\to\tau\nu_{\tau}$ & $1.04\pm0.15$ 
&$1.65\pm0.34$
&$1.54^{+0.38+0.29}_{-0.37-0.31}$&$1.83^{+0.53}_{-0.49}\pm0.24$\\
$B\to\mu\nu_{\mu}$ & $0.0046\pm0.0007$&$<0.01$\\
$B_c\to\tau\nu_{\tau}$ & $178\pm22$\\
$B_c\to\mu\nu_{\mu}$ & $0.73\pm0.09$\\
\end{tabular}
\end{ruledtabular}
\end{table}

In Table~\ref{tab:dbl} we also give predictions for the leptonic
decays of the $B_c$ meson. For the evaluation we use the CKM matrix
element $|V_{cb}|=0.039\pm0.0015$, obtained above, and the value of the
$B_c$ decay constant $f_{B_c}=(433\pm5)$~MeV calculated in our model
\cite{bcprop}. This value is in good agreement with the recent
lattice calculation $f_{B_c}^{\rm lattice}=0.427(5)$~GeV \cite{mdfhl}. We see
that the predicted $BR(B_c\to\tau\nu_{\tau})$ is of order of few
percent and, in principle, can be measured at LHC where $B_c$ mesons are
copiously produced. It is important to check this prediction
experimentally.   

In summary, we calculated branching fractions of the semileptonic and
leptonic $B$ meson decays in the framework of the relativistic quark
model, paying particular attention to the decays
involving  $\tau$ lepton. The obtained results are compared to the
recent experimental data. We find that for the decay branching
fractions reasonable agreement between the theory and
experiment is observed for the CKM matrix elements
$|V_{cb}|=0.039\pm0.0015$ and
$|V_{ub}|=(4.05\pm0.20)\times10^{-3}$. The largest deviations of our
theoretical predictions from experimental data occur for the ratios
$R(D)$ and $R(D^*)$, and they are about  1.75$\sigma$  and
2.4$\sigma$, respectively.  

The authors are grateful to D. Ebert, V. A. Matveev,
M. M\"uller-Preussker and V. I. Savrin  
for  useful discussions.
This work was supported in part by the {\it Russian
Foundation for Basic Research} under Grant No.12-02-00053-a.

\end{document}